# ARTICLE

# Superconductivity in hydrated $Li_x(H_2O)_y TaS_2$

Huanlong Liu,[a] Shangxiong Huangfu,[b] Hai Lin,[a] Xiaofu Zhang*[c,d] and Andreas Schilling*[a]

We have systematically studied the structural and physical properties of the superconducting hydrated $Li_x(H_2O)_y TaS_2$ ($0.22 \leq x \leq 0.58$, $y \approx 0.86$). The powder X-ray diffraction patterns suggest that all the samples are single-phase compounds, and the crystal structure is similar to that of $2H$-$TaS_2$ ($P6_3/mmc$). The transition temperature to superconductivity shows a dome-shape dependence on the lithium content $x$ with a maximum $T_c$ of 4.6 K for $x \approx 0.42$, which is larger than in corresponding optimally doped $2H$-$TaS_2$ superconductors without water or organic intercalants ($T_c \sim 4.2$ K). There are no signs for a charge-density-wave formation in hydrated $Li_x(H_2O)_y TaS_2$. While our magnetic data indicate a rather strongly type-II behavior, heat-capacity measurements reveal, like in other $2H$-$TaS_2$-type compounds, a reduced discontinuity $\Delta C_e/\gamma T_c \approx 0.8$ at $T_c$, which is smaller than the standard BCS value 1.43. From the corresponding Sommerfeld constants $\gamma$ and Debye temperatures $\Theta_D$ we can derive the parameter describing the electron-phonon coupling $\lambda_{ep}$ and the electron density of states $DOS(E_F)$ at the Fermi level as functions of $x$. While the variation of the $DOS(E_F)$ is consistent with that of $T_c$, indicating that the lithium intercalation is tuning $T_c$ via changing the $DOS(E_F)$ in $2H$-$Li_x(H_2O)_y TaS_2$, the simultaneous changes of $\lambda_{ep}$ and $\Theta_D$ may also play a certain role.

## Introduction

Superconductivity in layered transition-metal dichalcogenides (TMDs) has attracted increasing interest due to its relevance for the understanding of high-temperature superconductors, which encompass various quantum states including metallic, Mott insulating, and a state with charge-density waves (CDWs)[1-3]. $2H$-$TaS_2$ is a prominent representative member of the $2H$-TMDs family, in which both superconductivity and CDW exist. In the bulk $2H$-$TaS_2$, the incommensurate CDW order has a transition temperature $T_{CDW}$ at ~ 75 K, which coexists with superconductivity as the temperature is decreased below the superconducting transition temperature ($T_c$) of 0.8 K[4,5]. Similar to other TMDs, the enhancement of superconductivity in $2H$-$TaS_2$ is tuned mainly via chemical doping, intercalation, annealing, gating, and

a. Department of Physics, University of Zurich, Winterthurerstrasse 190, CH-8057 Zurich, Switzerland. E-mail: schilling@physik.uzh.ch
b. Laboratory for High Performance Ceramics, Empa, Überlandstrasse 129, CH-8600 Dübendorf, Switzerland
c. State Key Laboratory of Functional Materials for Informatics, Shanghai Institute of Microsystem and Information Technology, Chinese Academy of Sciences (CAS), Shanghai 200050, China. E-mail: zhangxf@mail.sim.ac.cn
d. 4CAS Center for Excellence in Superconducting Electronics, Shanghai 200050, China.

exfoliation or restacking[6-11]. The enhancement of superconductivity is usually accompanied by the suppression of CDWs. However, when reducing the thickness of 2$H$-TaS$_2$ to a few layers, the $T_c$ is found to increase with reducing the number of layers, contrary to the behavior of other TMDs[12]. Especially for the monolayer 2$H$-TaS$_2$, the $T_c$ reaches the highest value of 3.4 K as the CDW entirely disappears[13,14]. The enhancement of superconductivity, in addition to a possible commensurate CDW order[11], could be closely related to the enhanced electron-phonon coupling by reduced screening in few-layer TaS$_2$[13], in a similar way to the situation in oxygenated ultrathin 2$H$-TaS$_2$[10].

In the alkali-metal or transition-metal atom-intercalated TMDs, the enhancement of superconductivity has been well-understood[2,15-18]. The evolution of superconductivity can be explained by the suppression of CDW state, or the rigid-band model of the band structure. For inorganic metal-intercalated 2$H$-TaS$_2$, the $T_c$ of the optimally intercalated samples can generally reach the maximum as the CDW is sufficiently suppressed, then decrease with further increasing the intercalation level, thereby forming a dome-like $T_c$ dependence on the intercalation level (dome-shape), which is similar to the pressure dependence of $T_c$ seen in high-pressure experiments on 2$H$-TaS$_2$[9,19,20]. On the contrary, when an organic- or inorganic-chemical group intercalant enters the interlayer of 2$H$-TaS$_2$, the enhanced superconductivity is always accompanied by the complete disappearance of the CDW state[8,21,22].

It is very likely that the charge transfer from such intercalants to the Ta-S layer increases the electron density of states at the Fermi level (DOS($E_F$)). The $T_c$ of the hydrated phases $A_x$(H$_2$O)$_y$TaS$_2$ can be much higher than that with the same doping level in $A_x$TaS$_2$ ($A$ = Li, Na, K, Rb; $x \geq 0.33$)[23,24]. Taking lithium-intercalation as an example, the rigid-band model indicates that the intercalant does not significantly change the band structure of 2$H$-TaS$_2$[25]. The DOS has a peak at the Fermi level, and the Fermi energy is located nearly in the center of the conduction band, which is dominated by the half-filled Ta 5$d_{z^2}$ band [13,25-27]. This band structure is hardly changed at the CDW transition, or even in a thinned monolayer[13]. The Fermi level of hydrated phases $A_x$(H$_2$O)$_y$TaS$_2$ moves from the center to the near top of the $d_{z^2}$ band upon charge transfer from the intercalant to the conduction band, leading to a decrease of DOS($E_F$)[24,28], which should be detrimental to superconductivity. The above scenario based on the rigid-band theory apparently fails to account for the experimentally observed distinctly enhanced $T_c$ of the hydrates which are higher than that of the anhydrates. Undoubtedly, interlayered water molecules are favorable for superconductivity in 2$H$-TaS$_2$.

However, whether the enhancement of $T_c$ is due to the possible enlargement of the interlayer spacing, which in turn weakens the interlayer coupling of the Ta-S layers, or due to the shielding of the random Coulomb potential, is still a matter of debate[23,24,29-31]. To unravel this puzzle, we performed a systematic investigation on hydrated lithium-intercalated 2$H$-TaS$_2$ including transport, magnetization, and heat-capacity measurements. The CDW transition is found to be absent in all the investigated hydrated Li$_x$(H$_2$O)$_y$TaS$_2$ samples. We find that the $T_c$ firstly increases with $x$, and then decreases in 2$H$-Li$_x$(H$_2$O)$_y$TaS$_2$ with increasing hydrated lithium intercalation, forming a dome-shape with the maximum $T_c$ of at least 4.6 K. The variation of the DOS($E_F$) is consistent with that of $T_c$, indicating that the hydrated-lithium intercalation tunes $T_c$ via changing the DOS($E_F$) in 2$H$-Li$_x$(H$_2$O)$_y$TaS$_2$, although simultaneous changes of the electron-phonon coupling and phonon frequencies may also have to be taken into account.

## Results and discussion

In Fig. 1(a), we present the PXRD patterns at room temperature for all $Li_x(H_2O)_yTaS_2$ samples. All the samples have a similar crystal structure with the space group of $P6_3/mmc$. To more clearly reveal the structural differences between the $Li_x(H_2O)_yTaS_2$ compounds, we have performed a detailed refinement on all the PXRD patterns by using Fullprof[32]. The Rietveld refinement of the PXRD pattern of hexagonal $Li_{0.31}(H_2O)_yTaS_2$ is shown in Fig. 1(b)[33]. The relative coordinates in the Ta-S layer of the $Li_{0.31}(H_2O)_yTaS_2$ compound are the same as those of $2H$-$TaS_2$[34]. The Ta atoms bond with six surrounding S atoms to form a $[TaS_6]$ triangular prism, and a co-edged connection to construct the $TaS_2$ layer. The interlayer spacing of about 5.8 Å for the $Li_x(H_2O)_yTaS_2$ compounds is much larger than that of $2H$-$Li_xTaS_2$ (~3.4 Å)[35], which indicates the presence of interlayered water. The crystal structure with the possible position of the water molecule is qualitatively shown in the inset of Fig. 1(b), although the exact positions of the hydrogen and oxygen atoms are unknown. The lithium contents of these hydrated $Li_x(H_2O)_yTaS_2$ powders have been measured by inductively-coupled plasma mass spectrometry. The refinement results of the cell parameters are displayed in Fig. 1(c). The lattice parameter $c$ gradually decreases with lithium content increasing when compared with $2H$-$TaS_2$, and shows the significant change for $x > 0.42$. We observe a linear trend of a slight expansion of the Ta-S layer and a simultaneous contraction of interlayer spacing with increasing the lithium content, which is similar to $2H$-$Li_xTaS_2$[35], and is another manifestation of the evolution of the lithium content in $Li_x(H_2O)_yTaS_2$ samples. The thermogravimetric analysis of $Li_x(H_2O)_yTaS_2$ ($x = 0.27, 0.33$) samples in an argon atmosphere (Fig. 1(d)) showed that the weight of the samples decreases, and then stabilizes with increasing temperature. The interlayered crystal water content is essentially constant, namely 0.86 per formula unit, despite the differences in lithium content in the $Li_x(H_2O)_yTaS_2$ samples, and is also close to the reported results for β-$Na_{1/3}(H_2O)_{0.77-0.87}TaS_2$ [36,37]. In principle, the intercalated water content can be related not only to the lithium content but also to air humidity[37,38]. However, for the present conditions and when the lithium content is below $x \approx 0.08$, any absorbed water must enter the Ta-S layers on interstitial sites, as it does not qualitatively affect the crystal structure. For higher values of $x$, we assume that one or two $H_2O$ interlayers exist. Preliminary data indicate the presence of one $H_2O$ interlayer for $0.08 < x < 0.22$, and of possibly two $H_2O$ interlayers for higher lithium contents.

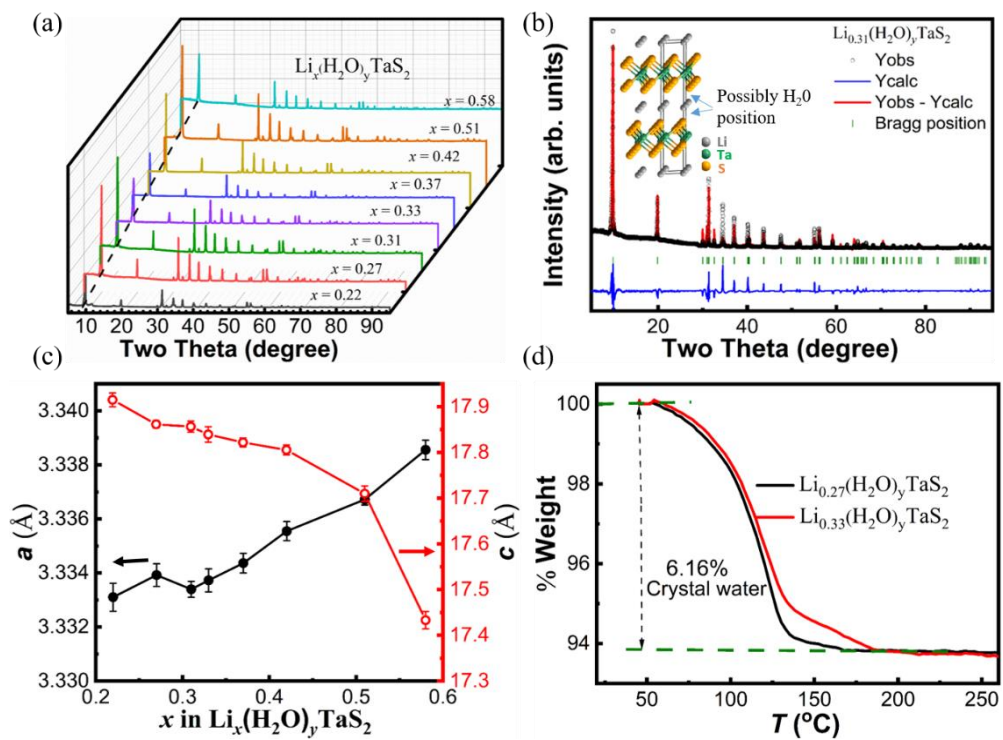

Fig. 1. (a) The PXRD pattern at ambient temperature for all samples of $Li_x(H_2O)_yTaS_2$ ($0.22 \leq x \leq 0.58$). (b) The PXRD pattern of $Li_{0.31}(H_2O)_yTaS_2$. The black dots are the observed data, while the red solid line represents the calculated intensities. The bottom blue solid line is the difference between the observed and calculated intensities. Insert: Crystal structure of $Li_{0.31}(H_2O)_yTaS_2$ along the ab-plane. (c) The variation of the cell parameters for $Li_x(H_2O)_yTaS_2$ samples. (d) The weight loss of $Li_x(H_2O)_yTaS_2$ ($x = 0.27$ and $0.33$) samples versus temperature in an argon gas environment.

Table 1. Fractional atomic and occupancy factors for $Li_{0.31}(H_2O)_yTaS_2$ ($R_{wp}$ = 8.047%, $\chi^2$ = 2.30). For this fit, the positions and occupancies of the light H and O atoms were omitted.

| Atom | Wyckoff position | x | y | z | Occ. |
|---|---|---|---|---|---|
| Ta | 2b | 0 | 0 | 0.2500 | 1 |
| S | 4f | 0.3333 | 0.1667 | 0.6180(4) | 1 |
| Li | 2a | 0 | 0 | 0 | 0.31 |

To explore the effect of lithium-intercalation into the hydrated phase $Li_x(H_2O)_yTaS_2$ on superconductivity, we have measured the resistivities $\rho(T)$ at temperatures ranging from 1.8 to 300 K for all $Li_x(H_2O)_yTaS_2$ samples, as shown in Fig. 2(a). All the samples show a weakly metallic behavior, which is similar to that of the parent $2H$-$TaS_2$[4]. The typical CDW-like kink in the $\rho(T)$ dependence completely disappears in the normal state for all samples when compared with the CDW transition temperature of ~75 K for $2H$-$TaS_2$[14,39]. The disappearance of the CDW transition may be due to the large amounts of intercalated lithium[8,24,40]. However, the interlayer spacing is also largely expanded by the water intercalation, which reduces the dimensionality and tends to suppress the CDW as well[23,29,41]. The detailed normal-to-superconducting transitions are shown in Fig. 2(b). The transition temperatures were determined by a 50% criterion and, alternatively, by the onset of the drop from the normal-state trend lines ($T_c^{onset}$). They first increase from $T_c \approx 2.8$ K to $\approx 4.5$ K ($T_c^{onset} \approx 5.0$ K), and then fall to ≈ 2.1 K ($T_c^{onset}$ ≈ 2.9 K), forming a dome-shape behavior with increasing intercalated lithium content in hydrated $Li_x(H_2O)_yTaS_2$, which is similar to the case of several families of high- and low-temperature superconductors [31,42-44].

The temperature-dependent DC magnetic susceptibilities of hydrated $Li_x(H_2O)_yTaS_2$ are shown in Fig. 2(c), which are measured with an external magnetic field of 2 mT and using the ZFC (zero-field cooled) and FC (field cooled) procedures. The $T_c$ is determined from the FC magnetization by either defining the intersection of the extrapolated normal-state magnetic susceptibility with the steepest slope line of the superconductivity signal, or alternatively by the onset temperature of diamagnetism. The results from all magnetic susceptibility and resistivity data are shown in Fig. 2(d), all in good agreement with respective $T_c$ data from heat-capacity measurements to be discussed further below. Lithium-intercalation is therefore an effective way to controllably tune superconductivity in hydrated $Li_x(H_2O)_yTaS_2$.

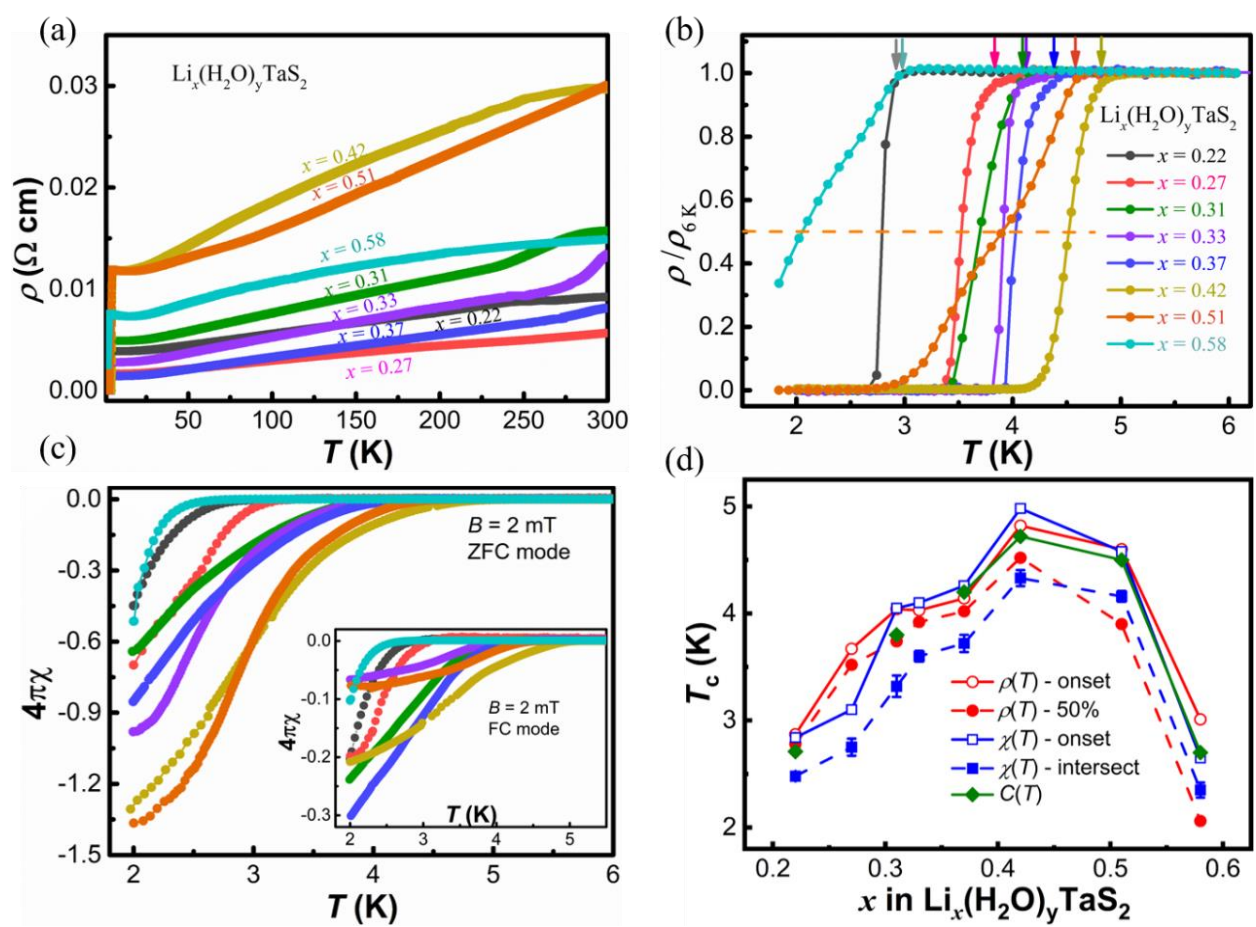

Fig. 2. The physical properties of $Li_x(H_2O)_yTaS_2$ samples. (a) Resistivity data in a temperature range between 1.8 and 300 K. (b) The resistivities $\rho/\rho_{6\,K}$ between 1.8 and 6 K, showing the transitions to superconductivity. The dashed line denotes the 50% criterion used to determine $T_c$, while the arrows indicate the onset temperatures $T_c^{onset}$ here the $\rho(T)$ curves start to drop from the normal-state trendlines. (c) The temperature-dependent magnetic susceptibilities for ZFC and FC procedures (inset), respectively. (d) The $T_c$ as determined by 50% of the normal resistivity ("$\rho(T)$-50%"), by the onset temperature of the resistivity drops ("$\rho(T)$-onset"), by corresponding data from FC susceptibility (steepest-slope method and onset temperature of diamagnetism, "$\chi(T)$-intersect" and "$\chi(T)$-onset", respectively), and from heat-capacity data ("$C(T)$").

ARTICLE

As an example, Figure 3(a) depicts the magneto-transport measurements for the optimally intercalated $Li_{0.42}(H_2O)_yTaS_2$. With gradually increasing the magnetic field to $B \leq 2.0$ T, the $\rho(T)$ curves show a systematic shift to low temperatures along with a certain field-induced broadening that is not uncommon in layered superconductors. For simplicity, we are using here the 50% criterion to evaluate the corresponding upper-critical fields $B_{c2}(T)$. A corresponding determination of $B_{c2}$ from magnetization $M(B)$ data as shown later in Fig. 4(b) is less definite because the transition is very broad in high magnetic fields, but it yields consistent $B_{c2}$ values (see below). The respective field-dependent critical temperatures for all investigated samples are summarized in Fig. 3(b). The field $B_{c2}(T)$ has a positive curvature as a function of temperature, which has also been found in other superconductors, such as organic intercalated-$TaS_2$, and $MgB_2$[8,45]. The polycrystalline average of the $B_{c2}(0)$ at zero temperature can be estimated with a fit according to[46]

$$B_{c2}(T) = B_{c2}(0)\left[1 - \left(\frac{T}{T_c}\right)^{3/2}\right]^{3/2} \quad (1),$$

The resulting $B_{c2}(0)$ values are higher than the reported upper-critical field 1.17 T for the pristine $2H$-$TaS_2$, but are still smaller than the corresponding Bardeen-Cooper-Schrieffer (BCS) weak-coupling Pauli limits (1.85$T_c$ in T).

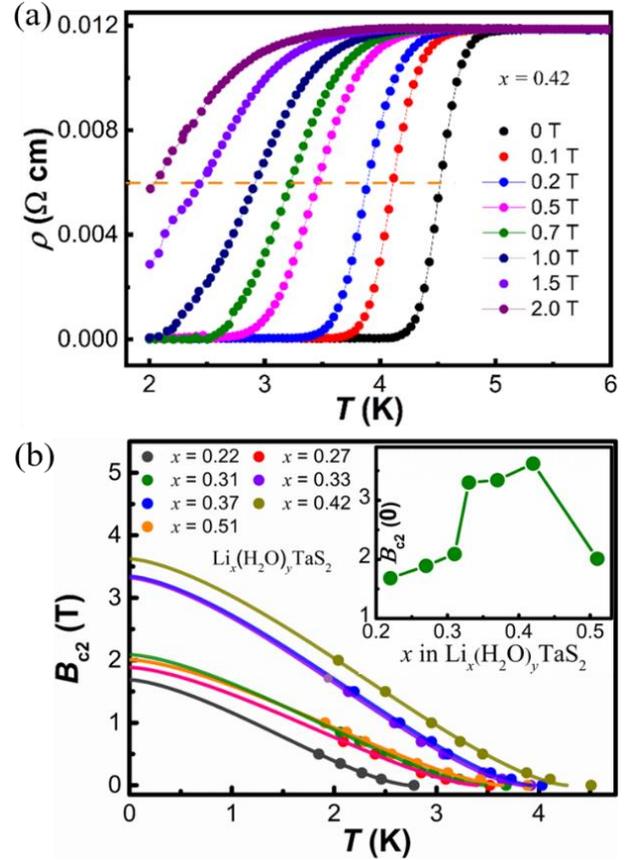

Fig. 3 (a) Temperature dependence of the resistivity of $Li_{0.42}(H_2O)_yTaS_2$ for various magnetic fields. The dashed line corresponds to the 50% criterion used to evaluate $B_{c2}$. (b) The estimation of the upper-critical field for $Li_x(H_2O)_yTaS_2$ (0.22 ≤ $x$ ≤ 0.51). The inset shows the respective extrapolated values of the upper-critical fields at $T = 0$.

In Fig. 4(a), we show the Meissner (FC) and the shielding (ZFC) signal of the optimally intercalated sample with $x$ = 0.42, measured in 2 mT. From the shielding fraction of 131.7% at 1.8 K obtained in the ZFC experiment, we estimate a demagnetization factor $N = 0.240(8)$ for this sample assuming an effective 100% shielding, and we used this value to correct all the magnetic low-field data, and all the parameters derived from them, according to $B_{\text{eff.}} = B - N * \mu_0 M$. While the critical temperature as determined from the DC magnetization, $T_c \approx$ 4.3 K - 5 K, is consistent with $T_c$ from resistivity data ($\approx$ 4.5 K - 4.8 K) (Fig. 2(b)), the calculated large effective Meissner volume of ~ 26% even in the FC experiment indicates significant magnetic-flux expulsion and the good quality of the sample. The magnetization loop of $Li_{0.42}(H_2O)_yTaS_2$ at 1.8 K shows a typically type-II behavior [Fig. 4(b) and upper inset therein]. In the lower inset of Fig. 4(b) we show a further estimate of $B_{c2}$ based on the first deviation of the magnetization from linearity, yielding $B_{c2} = 1.9 \pm 0.2$ T at 1.8 K, which is similar to the result of our magneto-transport data. In Fig 4(c), we plot the field dependence of the ZFC magnetization in the temperatures range between 1.8 and 4.6 K (in 0.2 K steps) for $Li_{0.42}(H_2O)_yTaS_2$. The dashed line shows the ideal linear behavior of the Meissner state ($\chi$ = -1). To obtain an estimate for the polycrystalline average of the lower critical fields $B_{c1}$, one can determine the $B_{c1}$ by the magnetic fields where the $M(B_{\text{eff.}})$ curves in Fig. 4(c) first deviate from linearity at different temperatures (here referred to as "method 1"). An improved approach, incorporating concepts of the Bean critical-state model for the mixed state and making corresponding assumptions on the mechanism of magnetic-flux penetration allows one to quantitatively fit the deviation δ$M(B_{\text{eff.}})$ from a linear $M(B_{\text{eff.}})$ to extract $B_{c1}$ (Ref. 47, here referred to as "method 2"). Lastly, a work including in addition the effect of Bean-Livingston barriers (Ref. 48, "method 3") suggested an improved fitting procedure of δ$M(B_{\text{eff.}})$ to eliminate this effect, at least at low temperatures. The resulting $B_{c1}$ values obtained by methods 1 and 2 are shown in Fig. 4(d), while our attempts to fit the δ$M(B_{\text{eff.}})$ data according to method 3 (which aimed to include edge-barrier effects) did not result in physically reasonable quantities. In this way we can still derive corresponding upper limits of $B_{c1}(0)$ from the empirical formula

$$B_{c1}(T) = B_{c1}(0)\left[1 - \left(\frac{T}{T_c}\right)^2\right], \qquad (2)$$

with $B_{c1}(0) \approx$ 2.7 mT (method 1, purple dashed line) and 2.1 mT (method 2, red dashed line), which we are using for the further analysis. To be on the safe side, however, these $B_{c1}(T)$ values should be interpreted as fields for the first magnetic-flux entry, which could indeed be larger than the true $B_{c1}$.

Together with $B_{c1}(0) = \frac{\Phi_0}{4\pi\lambda^2}\ln\frac{\lambda(0)}{\xi(0)}$ and $B_{c2} = \Phi_0/2\pi\xi(0)^2$, both the superconducting coherence length $\xi(0)$ and London penetration depth $\lambda(0)$ can be estimated to $\approx$ 9.6 nm and $\geq$ 69 nm at zero temperature, respectively. The resulting Ginzburg-Landau parameter $\kappa = \frac{\lambda(0)}{\xi(0)} \geq$ 7, indicates a type-II behavior of $Li_{0.42}(H_2O)_yTaS_2$. As the given $B_{c1}(0)$ and $B_{c2}(0)$ stem from extrapolations over an extended temperature region, the corresponding errors in these values and the quantities derived from them may be considerable, but they should reflect the correct orders of magnitude.

ARTICLE

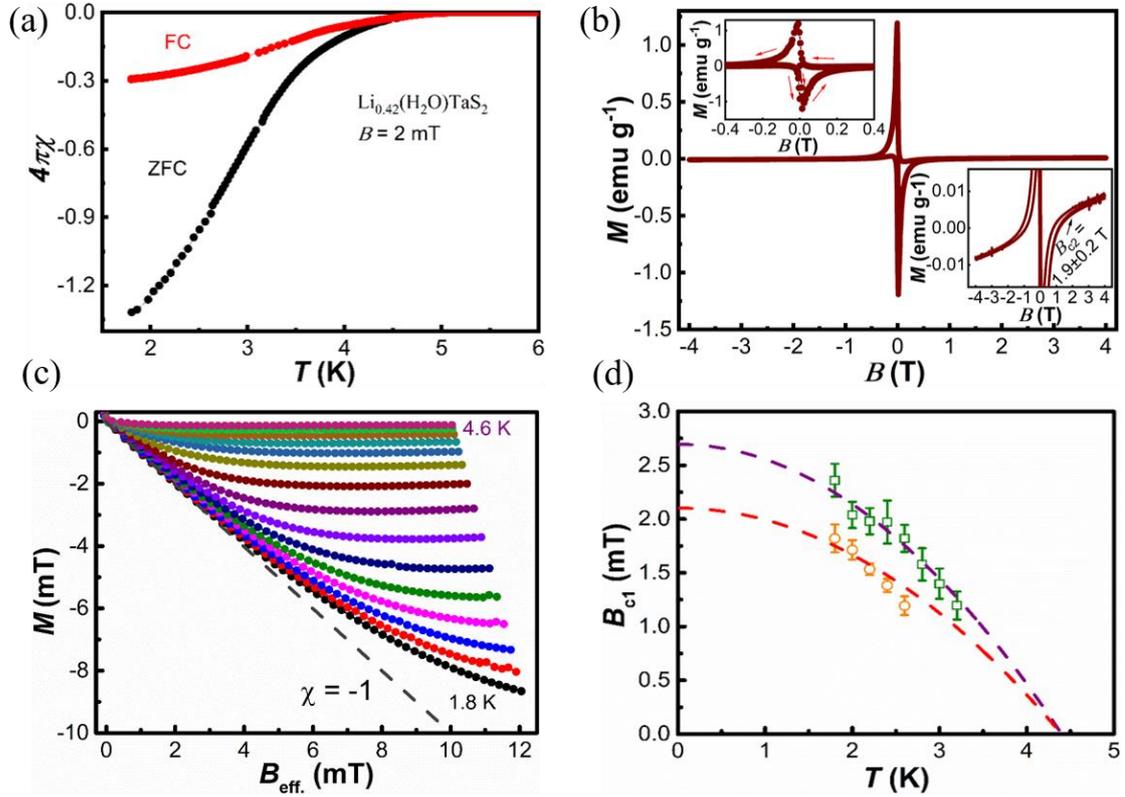

Fig 4. (a) The ZFC and FC magnetic susceptibilities, and (b) measured $M(B)$ loop at 1.8 K for optimally intercalated $Li_{0.42}(H_2O)_yTaS_2$. Insets: magnified views to show details. (c) The ZFC field dependence of the magnetization between temperatures 1.8 and 4.6 K (in 0.2 K steps) for $Li_{0.42}(H_2O)_yTaS_2$. The gray dashed line shows the ideal diamagnetic shielding. (d) The evaluated $B_{c1}$ values from Figs. 4(c) at different temperatures by methods 1 (squares) and 2 (circles) with the uncertainties from the extraction procedure drawn as error bars. The dashed lines are the fitting results using Eq.(2).

To further investigate the superconductivity in hydrated $Li_x(H_2O)_yTaS_2$, the temperature dependence of the heat capacity $C$ in zero magnetic field was measured from 1.9 to 10 K (Fig. 5(a), drawn as $C/T$). The $T_c$ from an entropy-conserving construction with an idealized specific heat discontinuity is $T_c \approx 4.6 - 4.7$ K as illustrated in the inset of Fig. 5(a), which is consistent with the magnetic susceptibility and resistivity data for $x = 0.42$ shown in Fig. 2(d). The normal-state contribution can be fitted using the data between $T_c$ to 10 K to

$$C/T = \gamma + \beta T^2, \qquad (3)$$

where the electronic contribution is $C_e = \gamma T$ and the phonon contribution is $C_{ph} = \beta T^3$. The Sommerfeld constant $\gamma$ is related to the DOS($E_F$) to be discussed further below, and $\beta$ is associated with the Debye temperature $\Theta_D$ via $\beta = 12\pi^4 nR/5\Theta_D^3$, with $n$ the number of atoms per formula unit and $R$ the gas constant. By fitting the specific-heat data to the above equation, we obtain the parameters summarized in Fig. 5(b). The fitted $\beta$ values at first increase and then decrease with increasing lithium content, resulting in a minimum of $\Theta_D$ of ~269 K, which is somewhat larger than in metal-intercalated 2$H$-TaS$_2$ and parent 2$H$-TaS$_2$ ($\Theta_D \approx 250$ K)[9,16,35,49]. It has been reported that in the organic-matter intercalated 2$H$-TaS$_2$, the $\Theta_D$ values are also larger than in the parent 2$H$-TaS$_2$. However, the reduction of $\Theta_D$ upon lithium intercalation in our samples by ~8.6% (Fig. 5(c)) may indicate a certain successive phonon softening associated with lithium doping. This is completely different from metal-intercalated 2$H$-TaS$_2$, where the corresponding quantities values hardly change with the intercalation[9,16,40].

By subtracting the phonon contribution from the total specific heats, the electronic parts of specific heat that characterize the superconducting state can be obtained, as it is shown for optimally hydrated Li$_{0.42}$(H$_2$O)$_y$TaS$_2$ with a maximum of $\Delta C_e/\gamma T_c \approx 0.81(5)$ in the inset of Fig. 5(a). This value is significantly lower than the expected standard BCS weak-coupling value of 1.43 and is similar to those reported for the parent 2$H$-TaS$_2$ and organic-matter intercalated 2$H$-TaS$_2$[29]. Therefore, the superconductivity in 2$H$-TaS$_2$ related compounds cannot be described by the BCS theory in its simplest form. The reduced discontinuity of the specific heat as compared to that of a single-band s-wave superconductor might indicate unconventional superconductivity with gap nodes or multiband features[50,51], or possible anisotropic single-gap superconductivity[52].

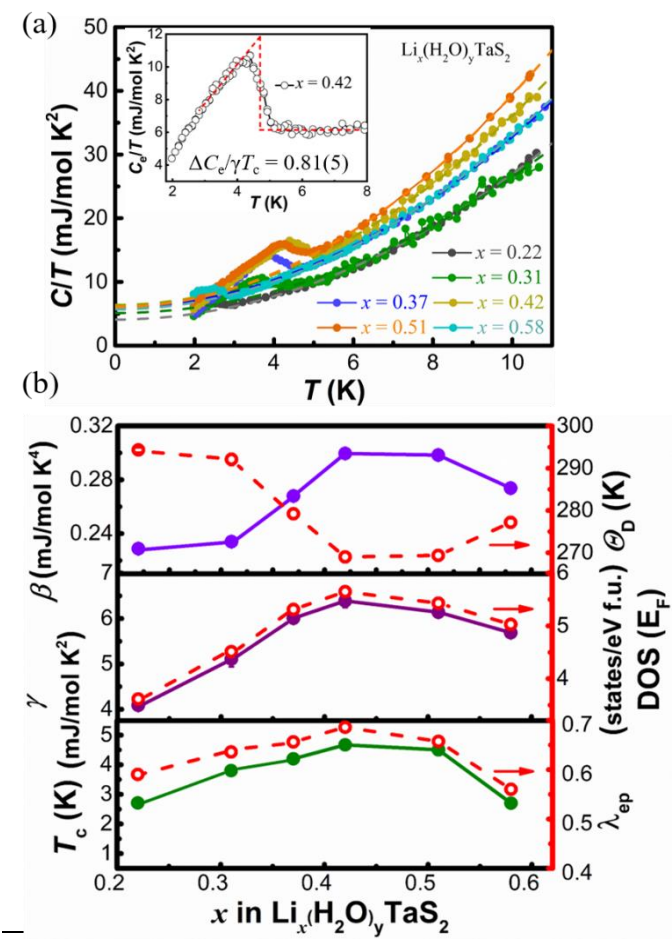

Fig. 5. (a) Reduced specific heat $C/T$ vs. $T$ curves of polycrystalline Li$_x$(H$_2$O)$_y$TaS$_2$ ($x$ = 0.22, 0.31, 0.37, 0.42, 0.51, and 0.58) samples. The dashed lines represent fits to the data between the superconducting transition temperature and 10 K. The inset shows the electronic specific heat $\Delta C_e / T$ at optimal intercalation with $x$ = 0.42, together with an entropy-conserving construction to evaluate the discontinuity in $C_e / T$. The red dashed line is a fit to the superconducting part with a single gap BCS model. (b) Variation of $\beta$ and corresponding $\Theta_D$, (c) normal state $\gamma$ and corresponding DOS($E_F$), and (d) the $T_c$ determined from the discontinuity in the specific heat, together with the estimated $\lambda_{ep}$, all as functions of lithium content $x$.

With the measured parameters $\Theta_D$ and $T_c$, the parameter describing the electron-phonon coupling contribution, $\lambda_{ep} = \text{DOS}(E_F)V_{\text{e-ph}}$ (where $V_{\text{e-ph}}$ denotes the electron-phonon-induced interaction), can be estimated from the inverted McMillan equation, assuming a repulsive screened-Coulomb parameter $\mu^* \approx 0.13$, which is a typical value for a metal system [53],

$$\lambda_{ep} = \frac{1.04+\mu^*\ln(\frac{\Theta_D}{1.45T_C})}{(1-0.62\mu^*)\ln(\frac{\Theta_D}{1.45T_C})-1.04}. \quad (4)$$

The estimated $\lambda_{ep} \approx 0.6 - 0.69$ (Fig. 5(d)) are larger than for parent $2H$-TaS$_2$ (~ 0.49) and are similar to those of the metal- intercalated $2H$-TaS$_2$, such as $2H$-Li$_x$TaS$_2$[35], $2H$-Ni$_x$TaS$_2$[9], $2H$-Cu$_x$TaS$_2$[16], and $2H$-TaS$_{2-x}$Se$_x$[54]. The enhancement of $T_c$ is reminiscent to the effect of the oxygenation of ultrathin $2H$-TaS$_2$, which has been attributed to an enhancement of the DOS($E_F$) by incorporation of oxygen into the TaS$_2$ crystal lattice, resulting in a strongly increased electron–phonon coupling[10]. It is worth noting that the $T_c$ of $\approx 4.6$ K (with an onset of diamagnetism up to $\approx 5.0$ K) for hydrated Li$_{0.42}$(H$_2$O)$_y$TaS$_2$ is somewhat larger than the value reported for optimally metal-intercalated $2H$-TaS$_2$ superconductors ($T_c \sim 4.2$ K), indicating that the presence of interlayered water is favorable to superconductivity, which is similar to the role of other interlayer organic matter in $2H$-TaS$_2$[55]. Using the $\lambda_{ep}$ and $\gamma$ values in appropriate units, the DOS($E_F$) can be estimated from the equation[56]

$$\text{DOS}(E_F) = \frac{3\gamma}{\pi^2 k_B^2(1+\lambda_{ep})}, \quad (5)$$

with the Boltzmann constant $k_B$. The calculated DOS($E_F$) values shown in Fig. 5(c) exhibit an increasing trend at first and then slightly decrease with increasing $x$, with the maximum of 5.6(5) states eV$^{-1}$ f.u.$^{-1}$ as the lithium content reaches $x = 0.42$, which is very similar to the associated variation of $T_c$ with $x$. The intercalation of organic matter has been reported to have no obvious effect on the DOS($E_F$) of the $2H$-TaS$_2$ host, but mainly to affect the interlayer spacing in the $2H$-TaS$_2$ system [55,57]. As $y \approx 0.86$ is constant for our $x \leq 0.42$ samples, the observed increase in the DOS($E_F$) must therefore be due to the intercalation of lithium only. The reason for the subsequent slight decrease of the DOS($E_F$) beyond $x \approx 0.42$ is unclear, but it is conceivable that water intercalation and/or electron doping may further change the size or shape of the Fermi surface as it has been found for the related $2H$-Cu$_x$TaS$_2$[16,24].

To gain some more insight into the enhanced superconductivity in Li$_x$(H$_2$O)$_y$TaS$_2$ based on our experimental results, we now focus on the $T_c$ given by the BCS theory [58] in terms of the DOS($E_F$), the total electron-electron interaction $V$, and $\Theta_D$,

$$T_c \sim \Theta_D \exp[-\frac{1}{\text{DOS}(E_F)V}], \quad (6)$$

The $T_c$ is proportional to $\Theta_D$ and is a strongly varying function of the density of mobile charge carriers since the DOS($E_F$) enters exponentially. As the calculated DOS($E_F$) values and $T_c$ follow a similar trendline upon lithium intercalation, it is tempting to make the variation of the DOS($E_F$) solely responsible for the dome-like $T_c$ dependence. However, the presence of the interlayered lithium atoms, together with the interlayer water, is also related to the variation of $\Theta_D$ (Fig. 5(b)) and $\lambda_{ep}$ (Fig. 5(d)), and hence, to the critical temperature $T_c$ via the effect on the phonon spectrum and the electron-phonon coupling. It is very interesting to note that a strengthening of the electron-phonon coupling and a certain simultaneous phonon softening have been observed in other systems in the context of a dome-shaped $T_c$ dependence [53,59,60].

Finally, the role of the interlayered water seems to be mainly to expand the interlayer spacing to ~0.58 nm for hydrated $Li_{0.42}(H_2O)_yTaS_2$, which is larger than that of parent $2H$-$TaS_2$ (~0.34 nm). Although this enlarged interlayer spacing does not seem to strongly correlate with superconductivity in our system, the situation of interlayered species in $2H$-$TaS_2$ can be complicated[29], and it is possible in principle that a weakened interlayer coupling and enhanced two-dimensionality can boost superconductivity as well [12,23,61].

## Conclusions

Hydrated $Li_x(H_2O)_yTaS_2$ ($0.22 \leq x \leq 0.58$; $y \approx 0.86$) shows a dome-shape dependence of the critical temperature $T_c$ on the lithium content $x$ with a maximum $T_c$ of at least 4.6 K for $x \approx 0.42$. This value is larger than in corresponding optimally intercalated $2H$-$TaS_2$ without water or organic intercalants, supporting the scenario that a weakened interlayer coupling - as a result of a large interlayer spacing - may suppress the tendency of charge-density-wave formation and enhance superconductivity. The electron density of states at the Fermi level $DOS(E_F)$ strongly varies with lithium content $x$ and closely follows the corresponding variation of the critical temperature. While this may be the main factor influencing $T_c$, we observe a simultaneous strengthening of the electron-phonon coupling and a phonon softening upon approaching the maximum critical temperature, which has also been reported to occur in other systems in the context of a dome-shaped $T_c$ dependence.

## Experimental methods

### Sample preparation

The $Li_x(H_2O)_yTaS_2$ samples were prepared by a reaction between $Li_xTaS_2$ and water molecules in air. The $Li_xTaS_2$ powders were obtained from 99.9% $Li_2S$, 99.99% Ta, and 99.9% S powders by solid-state reaction[35]. The raw materials were mixed, ground, pressed into tablets in a glove box, and then sealed into evacuated silica tubes. The tubes were then annealed at 800 °C for 12h in a muffle furnace. Finally, they were cooled down to room temperature along with the cooling of the furnace. The obtained $2H$-$Li_xTaS_2$ samples were exposed to air (relative humidity of ~31%) to form $Li_x(H_2O)_yTaS_2$, and then kept in a glove box with filled Argon gas.

### Experimental characterization

The structure was determined by powder-X-ray diffraction (PXRD) using a Stoe STADIP diffractometer at room temperature with Cu $K_{\alpha1}$ radiation. The lithium contents were measured by inductively coupled plasma mass spectrometry (ICP-MS) with an Agilent QQQ 8800 Triple quad ICP-MS spectrometer. The water content $y$ was determined by thermogravimetric analysis in argon atmosphere. The transport measurements and specific heats were carried out with a Physical Property Measurement System (PPMS, Quantum Design Inc.). The magnetic properties were studied in a Magnetic Properties Measurement System (MPMS3 from Quantum Design Inc.)

## Conflicts of interest

There are no conflicts to declare.

## Acknowledgements

This work was supported by the Swiss National Foundation under Grants No. 20-175554, 206021-150784.

## References

1. Saito, Y.; Nojima, T.; Iwasa, Y., Highly Crystalline 2d Superconductors. Nat. Rev. Mater., 2016, 2, 16094.

2. Chikina, A.; Fedorov, A.; Bhoi, D.; Voroshnin, V.; Haubold, E.; Kushnirenko, Y.; Kim, K. H.; Borisenko, S., Turning Charge-Density Waves into Cooper Pairs. npj Quantum Mater., 2020, 5, 22.

3. Woollam, J. A.; Somoano, R. B., Superconducting Critical Fields of Alkali and Alkaline-Earth Intercalates of $MoS_2$. Phys. Rev. B, 1976, 13, 3843-3853.

4. Nagata, S.; Aochi, T.; Abe, T.; Ebisu, S.; Hagino, T.; Seki, Y.; Tsutsumi, K., Superconductivity in the Layered Compound $2H-TaS_2$. Phys. Chem. Solids., 1992, 53, 1259-1263.

5. Omloo, W. P. F. A. M.; Jellinek, F., Intercalation Compounds of Alkali Metals with Niobium and Tantalum Dichalcogenides. J. less-common met., 1970, 20, 121-129.

6. J. Pan, C. Guo, C. Song, X. Lai, H. Li, W. Zhao, H. Zhang, G. Mu, K. Bu, T. Lin, X. Xie, M. Chen, and F. Huang, Enhanced Superconductivity in Restacked $TaS_2$ Nanosheets. J. Am. Chem. Soc. 2017, 139, 4623-4626.

7. Zhu, X. D.; Sun, Y. P.; Zhu, X. B.; Luo, X.; Wang, B. S.; Li, G.; Yang, Z. R.; Song, W. H.; Dai, J. M., Single Crystal Growth and Characterizations of $Cu_{0.03}TaS_2$ Superconductors. J. Cryst. Growth, 2008, 311, 218-221.

8. Wang, N. Z.; Shi, M. Z.; Shang, C.; Meng, F. B.; Ma, L. K.; Luo, X. G.; Chen, X. H., Tunable Superconductivity by Electrochemical Intercalation in $TaS_2$. New J. Phys., 2018, 20, 023014.

9. Li, L. J.; Zhu, X. D.; Sun, Y. P.; Lei, H. C.; Wang, B. S.; Zhang, S. B.; Zhu, X. B.; Yang, Z. R.; Song, W. H., Superconductivity of Ni-Doping $2H–TaS_2$. Physica C: Superconductivity 2010, 470, 313-317.

10. Achari, Amritroop; Bekaert, Jonas; Sreepal, Vishnu; Orekhov, Andrey; Kumaravadivel, Piranavan; Kim, Minsoo; Gauquelin, Nicolas; Balakrishna Pillai, Premlal; Verbeeck, Johan; Peeters, Francois M; Geim, Andre K; Milošević, Milorad V; Nair, Rahul R. Alternating Superconducting and Charge Density Wave Monolayers within Bulk $6R-TaS_2$. Nano Lett 2022, 22, 6268-6275.

11. Jonas Bekaert, Ekaterina Khestanova, David G. Hopkinson, John Birkbeck, Nick Clark, Mengjian Zhu, Denis A. Bandurin, Roman Gorbachev, Simon Fairclough, Yichao Zou, Matthew Hamer, Daniel J. Terry, Jonathan J. P. Peters, Ana M. Sanchez, Bart Partoens, Sarah J. Haigh, Milorad V. Milošević, and Irina V. Grigorieva. Enhanced Superconductivity in Few-Layer $TaS_2$ due to Healing by Oxygenation. Nano Lett 2020, 20, 3808-3818.

12. Yan, R.; Khalsa, G.; Schaefer, B. T.; Jarjour, A.; Rouvimov, S.; Nowack, K. C.; Xing, H. G.; Jena, D., Thickness Dependence of Superconductivity in Ultrathin $NbS_2$. Appl. Phys. Express, 2019, 12, 023008.

13. N. M. Efrén, I. Joshua O, M. V. Samuel, P. C. Elena, C. G. Andres, Q. Jorge, R. B. Gabino, C. Luca, S. G. Jose Angel, A. Nicolás, and S. Gary A, G. Francisco, van der Z. Herre S. J. and C. Eugenio, Enhanced Superconductivity in Atomically Thin $TaS_2$. Nat. Commun., 2016, 7, 11043.

14. Yang, Y.; Fang, S.; Fatemi, V.; Ruhman, J.; Navarro-Moratalla, E.; Watanabe, K.; Taniguchi, T.; Kaxiras, E.; Jarillo-Herrero, P., Enhanced Superconductivity Upon Weakening of Charge Density Wave Transport in $2H-TaS_2$ in the Two-Dimensional Limit. Phys. Rev. B, 2018, 98, 035203.

15. D. C. Freitas, P. Rodière, M. R. Osorio, E. Navarro-Moratalla, N. M. Nemes, V. G. Tissen, L. Cario, E. Coronado, M. García-Hernández, S. Vieira, M. Núñez-Regueiro, and H. Suderow, Strong Enhancement of Superconductivity at High Pressures within the Charge-Density-Wave States of $2H-TaS_2$ and $2H-TaSe_2$. Phys. Rev. B, 2016, 93, 184512.

16. K. E. Wagner, E. Morosan, Y. S. Hor, J. Tao, Y. Zhu, T. Sanders, T. M. McQueen, H. W. Zandbergen, A. J. Williams, D. V. West, and R. J. Cava, Tuning the Charge Density Wave and Superconductivity in $Cu_xTaS_2$. Phys. Rev. B, 2008, 78, 104520.

17. Zhao, X.-M.; Zhang, K.; Cao, Z.-Y.; Zhao, Z.-W.; Struzhkin, V. V.; Goncharov, A. F.; Wang, H.-K.; Gavriliuk, A. G.; Mao, H.-K.; Chen, X.-J., Pressure Tuning of the Charge Density Wave and Superconductivity in $2H-TaS_2$. Phys. Rev. B, 2020, 101, 134506.

18. Castro Neto, A. H., Charge Density Wave, Superconductivity, and Anomalous Metallic Behavior in 2d Transition Metal Dichalcogenides. Phys. Rev. Lett., 2001, 86, 4382-4385.

19. Peng, J.; Yu, Z.; Wu, J.; Zhou, Y.; Guo, Y.; Li, Z.; Zhao, J.; Wu, C.; Xie, Y., Disorder Enhanced Superconductivity toward $TaS_2$ Monolayer. ACS Nano, 2018, 12, 9461-9466.

20. S. X. Xu, J. J. Gao, Z. Y. Liu, K. Y. Chen, P. T. Yang, S. J. Tian, C. S. Gong, J. P. Sun, M. Q. Xue, J. Gouchi, X. Luo, Y. P. Sun, Y. Uwatoko, H. C. Lei, B. S. Wang, and J. G. Cheng, Effects of Disorder and Hydrostatic Pressure on Charge Density Wave and Superconductivity in $2H-TaS_2$. Phys. Rev. B, 2021, 103, 224509.

21. Gamble, F. R.; DiSalvo, F. J.; Klemm, R. A.; Geballe, T. H., Superconductivity in Layered Structure Organometallic Crystals. Science 1970, 168, 568.

22. Gamble, F. R.; Osiecki, J. H.; Cais, M.; Pisharody, R.; DiSalvo, F. J.; Geballe, T. H., Intercalation Complexes of Lewis Bases and Layered Sulfides: A Large Class of New Superconductors. Science 1971, 174, 493.

23. Katayama, N.; Nohara, M.; Sakai, F.; Takagi, H., Enhanced Superconducting Transition Temperature in the Water-Intercalated Sulfides. J. Phys. Soc. Japan., 2005, 74, 851-854.

24. Lerf, A.; Sernetz, F.; Biberacher, W.; Schöllhorn, R., Superconductivity in Layered Ternary Chalcogenides $A_xTaS_2$ and $A_xNbS_2$ and Influence of Topotactic Solvation. Mater. Res. Bull., 1979, 14, 797-805.

25. Fang, L.; Wang, Y.; Zou, P. Y.; Tang, L.; Xu, Z.; Chen, H.; Dong, C.; Shan, L.; Wen, H. H., Fabrication and Superconductivity of $Na_xTaS_2$ Crystals. Phys. Rev. B, 2005, 72, 014534.

26. Zhao, R.; Grisafe, B.; Ghosh, R. K.; Holoviak, S.; Wang, B.; Wang, K.; Briggs, N.; Haque, A.; Datta, S.; Robinson, J., Two-Dimensional Tantalum Disulfide: Controlling Structure and Properties Via Synthesis. 2D Mater., 2018, 5, 025001.

27. Guo, G.; Liang, W., Electronic Structures of Intercalation Complexes of the Layered Compound 2H-$TaS_2$. J. Phys. C, 1987, 20, 4315.

28. Biberacher, W.; Lerf, A.; Besenhard, J.; Möhwald, H.; Butz, T.; Saibene, S., Electrointercalation into 2H-$TaS_2$ Single Crystals: In Situ Dilatometry and Superconducting Properties. Il Nuovo Cimento D 1983, 2, 1706-1711.

29. Salvo, F. J. D.; Schwall, R.; Geballe, T. H.; Gamble, F. R.; Osiecki, J. H., Superconductivity in Layered Compounds with Variable Interlayer Spacings. Phys. Rev. Lett., 1971, 27, 310-313.

30. Thompson, A. H.; Gamble, F. R.; Koehler, R. F., Effects of Intercalation on Electron Transport in Tantalum Disulfide., Phys. Rev. B, 1972, 5, 2811-2816.

31. Sakurai, H.; Takada, K.; Izumi, F.; Dilanian, R. A.; Sasaki, T.; Takayama-Muromachi, E., The Role of the Water Molecules in Novel Superconductor, $Na_{0.35}CoO_2 \cdot 1.3H_2O$. Physica C: Superconductivity 2004, 412-414, 182-186.

32. Rodriguez-Carvajal, J., Collected Abstract of Powder Diffraction Meeting. Toulouse, France 1990, 127.

33. Mao, Y.; Fang, Y.; Pan, J.; Wang, D.; Bu, K.; Che, X.; Zhao, W.; Huang, F., Synthesis, Crystal Structures and Physical Properties of $A(H_2O) MoS_2$ (A = K, Rb, Cs). J. Solid State Chem., 2019, 279, 120937.

34. Meetsma, A.; Wiegers, G.; Haange, R.; De Boer, J., Structure of $2H-TaS_2$. Acta Crystallogr., Sect. C: Cryst. Struct. Commun., 1990, 46, 1598-1599.

35. Liu, H.; Huangfu, S.; Zhang, X.; Lin, H.; Schilling, A., Superconductivity and Charge Density Wave Formation in Lithium-Intercalated $2H-TaS_2$. Phys. Rev. B, 2021, 104, 064511.

36. Graf, H. A.; Lerf, A.; Schöllhorn, R., A Crystal Structure Investigation of the Hydrated Layered Chalcogenides $K_x(H_2O)_yNbS_2$ and $K_x(H_2O)_yTaS_2$. J. Less Common Met., 1977, 55, 213-220.

37. Johnston, D. C., Ambient Temperature Phase Relations in the System if $Na_{1/3}(H_2O)_yTaS_2$ (0<-y<-2). Mater. Res. Bull., 1982, 17, 13-23.

38. Johnston, D. C.; Frysinger, S. P., X-Ray Diffraction Study of $Na_{1/3}(H_2O)_{1.5}TaS_2$: Observation of a Hendricks-Teller Disordered Layer Lattice. Phys. Rev. B, 1984, 30, 980-984.

39. Hu, W. Z.; Li, G.; Yan, J.; Wen, H. H.; Wu, G.; Chen, X. H.; Wang, N. L., Optical Study of the Charge-Density-Wave Mechanism in $2H-TaS_2$ and $Na_xTaS_2$. Phys. Rev. B, 2007, 76, 045103.

40. Morosan, E.; Zandbergen, H. W.; Dennis, B. S.; Bos, J. W. G.; Onose, Y.; Klimczuk, T.; Ramirez, A. P.; Ong, N. P.; Cava, R. J., Superconductivity in $Cu_xTiSe_2$. Nat. Phys., 2006, 2, 544-550.

41. Gygax, S.; Biberacher, W.; Lerf, A.; Denhoff, M., Superconducting Parameters of a Strongly Anisotropic Intercalated $TaS_2$ Compound: $K_{0.33}(H_2O)_{0.66}TaS_2$. Helv. Phys. Acta, 1983, 55, 755-763.

42. J. Chang, E. Blackburn, A. T. Holmes, N. B. Christensen, J. Larsen, J. Mesot, R. Liang, D. A. Bonn, W. N. Hardy, A. Watenphul, M. v. Zimmermann, E. M. Forgan, and S. M. Hayden, Direct Observation of Competition between Superconductivity and Charge Density Wave Order in $YBa_2Cu_3O_{6.67}$. Nat. Phys., 2012, 8, 871-876.

43. B. J. Ramshaw, S. E. Sebastian, R. D. McDonald, J. Day, B. S. Tan, Z. Zhu, J. B. Betts, R. Liang, D. A. Bonn, W. N. Hardy, and N. Harrison, Quasiparticle Mass Enhancement Approaching Optimal Doping in a High-$T_c$ Superconductor. Science 2015, 348, 317.

44. Canfield, P. C.; Bud'ko, S. L., Feas-Based Superconductivity: A Case Study of the Effects of Transition Metal Doping on $BaFe_2As_2$. Annu. Rev. Condens. Matter Phys., 2010, 1, 27-50.

45. Müller, K.-H.; Fuchs, G.; Handstein, A.; Nenkov, K.; Narozhnyi, V.; Eckert, D., The Upper Critical Field in Superconducting MgB$_2$. J. Alloys Compd., 2001, 322, L10-L13.

46. Micnas, R.; Ranninger, J.; Robaszkiewicz, S., Superconductivity in Narrow-Band Systems with Local Nonretarded Attractive Interactions. Rev. Mod. Phys., 1990, 62, 113-171.

47. Moshchalkov, V. V., Henry, J. Y., Marin, C., Rossat-Mignod, J. & Jacquot, J. F. Anisotropy of the first critical field and critical current in YBa2Cu3O6.9 single crystals. Physica C, 1991, 175, 407-418.

48. Burlachkov, L., Yeshurun, Y., Konczykowski, M. & Holtzberg, F. Explanation for the low-temperature behavior of Hc1 in YBa$_2$Cu$_3$O$_7$. Phys. Rev. B, 1992, 45, 8193(R).

49. Abdel-Hafiez, M.; Zhao, X. M.; Kordyuk, A. A.; Fang, Y. W.; Pan, B.; He, Z.; Duan, C. G.; Zhao, J.; Chen, X. J., Enhancement of Superconductivity under Pressure and the Magnetic Phase Diagram of Tantalum Disulfide Single Crystals. Sci. Rep., 2016, 6, 31824.

50. M. Abdel-Hafiez, S. Aswartham, S. Wurmehl, V. Grinenko, C. Hess, S.-L. Drechsler, S. Johnston, A. U. B. Wolter, B. Büchner, H. Rosner, and L. Boeri, Specific Heat and Upper Critical Fields in KFe$_2$As$_2$ single Crystals. Phys. Rev. B, 2012, 85, 134533.

51. Milošević, M. V. & Perali, A. Emergent phenomena in multicomponent superconductivity: an introduction to the focus issue. Supercond. Sci. Technol. 2015, 28, 060201.

52. J. Bekaert, S. Vercauteren, A. Aperis, L. Komendova, R. Prozorov, B. Partoens, and M. V. Milosevic. Anisotropic type-I superconductivity and anomalous superfluid density in OsB$_2$. Phys. Rev. B, 2016, 94, 144506.

53. Kudo, K.; Takasuga, M.; Okamoto, Y.; Hiroi, Z.; Nohara, M., Giant Phonon Softening and Enhancement of Superconductivity by Phosphorus Doping of BaNi$_2$As$_2$. Phys. Rev. Lett., 2012, 109, 097002.

54. L. Li, X. Deng, Z.Wang, Y. Liu, M. Abeykoon, E. Dooryhee, A. Tomic, Y. Huang, J. B. Warren, E. S. Bozin, S. J. L. Billinge, Y. Sun, Y. Zhu, G. Kotliar, and C. Petrovic, Superconducting Order from Disorder in 2H-TaSe$_{2-x}$S$_x$. npj Quantum Mater., 2017, 2, 11.

55. Schlicht, A.; Schwenker, M.; Biberacher, W.; Lerf, A., Superconducting Transition Temperature of 2H−TaS$_2$ Intercalation Compounds Determined by the Phonon Spectrum. J. Phys. Chem. B, 2001, 105, 4867-4871.

56. Arushi; Singh, D.; Biswas, P. K.; Hillier, A. D.; Singh, R. P., Unconventional Superconducting Properties of Noncentrosymmetric Re$_{5.5}$Ta. Phys. Rev. B, 2020, 101, 144508.

57. Lerf, A., Different Modes and Consequences of Electron Transfer in Intercalation Compounds. J. Phys. Chem. Solids, 2004, 65, 553-563.

58. Bardeen, J.; Cooper, L. N.; Schrieffer, J. R., Theory of Superconductivity. Physical Review, 1957, 108, 1175-1204.

59. Flores-Livas, J. A.; Debord, R.; Botti, S.; San Miguel, A.; Marques, M. A.; Pailhes, S., Enhancing the Superconducting Transition Temperature of Basi2 by Structural Tuning. Phys. Rev. Lett., 2011, 106, 087002.

60. Guo, J.; Qi, Y.; Matsuishi, S.; Hosono, H., T C Maximum in Solid Solution of Pyrite Irse2–Rhse2 Induced by Destabilization of Anion Dimers. *J. Am. Chem. Soc.*, 2012, 134, 20001-20004.

61. Hinsche, N. F.; Thygesen, K. S., Electron–Phonon Interaction and Transport Properties of Metallic Bulk and Monolayer Transition Metal Dichalcogenide $TaS_2$. *2D Mater.*, 2017, 5, 015009.